# Enabling materials for subwavelength-size low-loss surface modes in the Terahertz spectral range

Mathieu Poulin, Steven Giannacopoulos and Maksim Skorobogatiy

*Ecole Polytechnique de Montréal, Department of Engineering Physics, C.P. 6079, Succ. Centre-Ville, Montreal, Québec, Canada H3C3A7*

## ABSTRACT

*Terahertz spectral range (frequencies of 0.1-1 THz) has recently emerged as the next frontier for non-destructive imaging, industrial sensing and ultra-fast wireless communications. Here, we review several classes of materials such as simple metals, semiconductors, high-k dielectrics, polar materials, zero gap materials, as well as structured materials that can support strongly localised electromagnetic modes at material interfaces in the Terahertz spectral range. We present the basic theory of surface waves, detail the requirement of strong modal confinement and low loss for the surface waves propagating at material interfaces and discuss challenges for excitation of such modes at Terahertz frequencies. A large number of examples related to naturally occurring and artificial materials is then presented. A variety of practical applications is envisioned for surface waves at Terahertz frequencies including non-destructive super-resolution imaging and quality control, high sensitivity sensors capable of operation with small volumes of analytes that are opaque in the visible and near-infrared, as well as design of compact optical circuit for the upcoming ultra-high bitrate THz communication devices.*

## INTRODUCTION

The extreme spatial confinement of the surface plasmon polaritons in the visible spectral range leads to a revolution in several application domains [1]. For example, plasmons play a major role in various surface sensing applications due to plasmon deeply sub-micron modal size [2, 3]. Additionally, plasmonic modes are promising for super-resolution imaging applications [4, 5]. Finally, plasmonic modes promise to bring great advances in integrated optical circuits for communications and other photonic applications as plasmon strong confinement allows to drastically reduce the size of photonic chips [6, 7]. Recently, THz spectral range (frequencies of 0.1-1 THz, wavelengths of 3mm-0.3mm) emerged as the next frontier for non-destructive imaging and ultra-fast wireless communications [8, 9]. Particularly, due to the relative transparency of dry dielectrics to THz radiation, many applications in non-destructive quality monitoring and packaging are envisioned. Moreover, the push to 5G and faster wireless networks demands development of the integrated optical circuits operating at THz frequencies. All these practical applications can benefit from tightly confined subwavelength modes like plasmons. Unfortunately, in the THz spectral range, generation of highly confined modes on metal surfaces proved to be challenging due to material limitations of metals, thus resulting in plasmon-polaritons that are highly delocalized and extend by 100-1000 wavelengths into the dielectric region.

The goal of the paper is to review alternative materials that could allow surface-bound electromagnetic modes with subwavelength confinement at THz frequencies. As argued in the paper, various semiconductors, polaritonic materials, zero-gap materials, high permittivity material and metamaterials can support deeply subwavelength THz surface states on their surfaces, thus enabling many of the same applications in imaging, communications and sensing that proved ground breaking when using surface plasmon-polaritons in the visible/near-IR.

## THEORY OF SURFACE WAVES

Interaction between electromagnetic (E&M) waves and free charge carriers can yield surface plasmon-polariton (SPP) excitation at the metal or semiconductor / dielectric interfaces, while interaction of E&M waves with lattice vibrational modes lead to surface phonon-polaritons at the polar media/dielectric interface. Moreover, well-confined surface-bound E&M states can be excited at the interface with a high-k dielectric having high material losses, which are known as a Zenneck surface waves. Finally, surface waves can be excited at the interface with a naturally occurring or artificially structured anisotropic media, in which case such waves are known as Dyakonov surface waves. Each one of these surface waves has its advantages and limitations and require a particular set of material properties for their excitation. Moreover, due to variation of the material properties with frequency, such surface waves can have completely different optical properties depending on the frequency range of operation. In this work we focus specifically on surface waves in the THz spectral range.

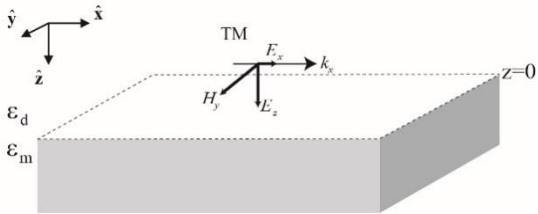

Figure 1. Schematic of an interface between two materials that guides a TM polarised surface wave.

We first review the necessary conditions for the excitation of surface waves at the interface between two distinct materials, one of them being lossless dielectric with positive dielectric permittivity $\varepsilon_d$ and another material with dielectric permittivity $\varepsilon_m$ that can have either negative real part (metal, semiconductor, polar media just above the resonance) or positive real part (polar media just below the resonance, high-k dielectric). Fig. 1 shows planar interface between two materials (OXY plane), and a TM-polarised electromagnetic (E&M) wave that has the only component of its magnetic field directed along the OY axis.

On either side of the interface we assume a single TM polarized planewave of frequency $\omega$ and the corresponding E&M wavevectors $\boldsymbol{k_{d,m}} = (k_x, 0, k_z^{d,m})$. The corresponding magnetic and electric fields on either side of the interface are given by:

$$\boldsymbol{H_{d,m}}(\boldsymbol{r},t) = \hat{y} H_{d,m} e^{i\boldsymbol{k_{d,m}}\boldsymbol{r}-i\omega t} \tag{1}$$

$$\boldsymbol{E_{d,m}}(\boldsymbol{r},t) = \frac{\boldsymbol{H_{d,m}}(\boldsymbol{r},t)\times \boldsymbol{k_{d,m}}}{\varepsilon_{d,m} k_0}, \tag{2}$$

where $k_0 = \omega/c = 2\pi/\lambda$, and the wavevectors satisfy the standard dispersion relations:

$$\boldsymbol{k_{d,m}^2} = \varepsilon_{d,m} k_0^2 = k_x^2 + \left(k_z^{d,m}\right)^2 \tag{3}$$

The E&M field components parallel to the interface must be continuous across the interface ($\hat{y}\boldsymbol{H_d}(0,t) = \hat{y}\boldsymbol{H_m}(0,t); \hat{x}E_d(0,t) = \hat{x}E_m(0,t)$), thus leading to the following equation for the $k_z^{d,m}$:

$$\frac{k_z^d}{\varepsilon_d} - \frac{k_z^m}{\varepsilon_m} = 0 \tag{4}$$

Using dispersion relations (3), as well as definition of the modal effective refractive index $k_x = n_{eff} k_0$, equation (4) admits a simple analytical solution for the surface mode dispersion relation:

$$\varepsilon_{eff} = n_{eff}^2 = \frac{\varepsilon_d \varepsilon_m}{\varepsilon_d + \varepsilon_m} \quad ; \quad \left(k_z^{d,m}\right)^2 = k_0^2 \frac{(\varepsilon_{d,m})^2}{\varepsilon_d + \varepsilon_m} \tag{5}$$

While (5) is a solution of (4), it does not necessarily describe a bound surface state as (4) is just a condition of E&M field continuity across the interface. In order for (5) to describe a surface state, one must require that the electromagnetic fields are decaying away from the interface (evanescent fields), which amounts to additional requirements:

$$\text{Im}(k_z^d) < 0 \quad ; \quad \text{Im}(k_z^m) > 0 \tag{6}$$

and define exponentially decaying fields in both media. Finally, for a surface bound state we define its propagation length $L_x$, as well as its extents into a dielectric $L_d$, and a second material $L_m$ as characteristic distances over which the E&M fields decay by the factor $e$:

$$L_x = \frac{1}{\text{Im}(k_x)} \quad ; \quad L_{d,m} = \frac{1}{\mp\text{Im}\left(k_z^{d,m}\right)} \tag{7}$$

Now we detail several types of surface states propagating at the interface between a classic lossless dielectric with $\varepsilon_d > 0$ and a second material that can be either a metal, a polar material, a zero-gap mterial, or a lossy high-k dielectric. Particularly, in case of simple Drude metals and semiconductors their frequency dependent dielectric constant is described as [10]:

$$\varepsilon_m(\omega) = 1 - \frac{\omega_p^2}{\omega^2 + i\gamma\omega}, \tag{8}$$

where $\omega_p$ is a plasma frequency and $\gamma$ is a damping coefficient. Similarly, dielectric constant of many polar materials near the resonance frequency $\omega_0$ can be described using the Lorentz model:

$$\varepsilon_m(\omega) = 1 - \frac{\omega_p^2}{\omega^2 - \omega_0^2 + i\gamma\omega}, \tag{9}$$

while lossy high-k dielectrics are characterised by $\varepsilon_m = \varepsilon_m' + i\varepsilon_m''$, where $\varepsilon_m' \gg 1$.

**Lossless materials**

We first consider the case of ideal (lossless) materials $\gamma = 0$ operating at frequencies where their dielectric constant is negative so that the two following conditions are satisfied $\varepsilon_m < 0$ and $\varepsilon_d + \varepsilon_m < 0$. In this case, the surface mode effective dielectric constant as given by (5) is purely real and positive $\varepsilon_{eff} > \varepsilon_d$. For metals and polar materials, this limits the operation frequency to:

$$\text{Metals: } \omega < \frac{\omega_p}{\sqrt{1+\varepsilon_d}} \quad ; \quad \text{Polar materials: } \omega_0 < \omega < \sqrt{\omega_0^2 + \frac{\omega_p^2}{1+\varepsilon_d}} \tag{10}$$

At such frequencies, the OZ components of the two wavevectors are purely imaginary and their imaginary parts have the opposite signs:

$$k_z^{d,m} = \mp i k_0 \frac{|\varepsilon_{d,m}|}{\sqrt{|\varepsilon_m| - \varepsilon_d}}, \tag{11}$$

thus defining a true surface state, which is called a surface plasmon-polariton in case of metals and surface phonon-polariton in case of polar materials. Electromagnetic fields of such a surface state decay exponentially fast away from the interface. The surface mode propagation length $L_x$ is infinite as $n_{eff}$ is purely real, while the surface mode extents into a dielectric $L_d$, and a material $L_m$ are given by (7):

$$L_{d,m} = \frac{\lambda}{2\pi} \frac{\sqrt{|\varepsilon_m| - \varepsilon_d}}{|\varepsilon_{d,m}|} \tag{12}$$

Note that plasmon-polariton or phonon-polariton extents into the two media become deeply subwavelength $L_{d,m} \ll \lambda$ when $|\varepsilon_m| \sim \varepsilon_d$, which happens near the following frequencies:

$$\text{Metals: } \omega \sim \frac{\omega_p}{\sqrt{1+\varepsilon_d}} \quad ; \quad \text{Polar materials: } \omega \sim \sqrt{\omega_0^2 + \frac{\omega_p^2}{1+\varepsilon_d}} \tag{13}$$

In figure 2 we present a typical loss-less surface plasmon-polariton dispersion relation as well as dependence of its penetration depths into a dielectric and metal as a function of frequency. The surface plasmon-polariton extent into metal is always deeply subwavelength $L_z^m \ll \lambda$, while its extent into dielectric is generally comparable or even smaller than the wavelength of light $L_z^d < \lambda$ at higher frequencies.

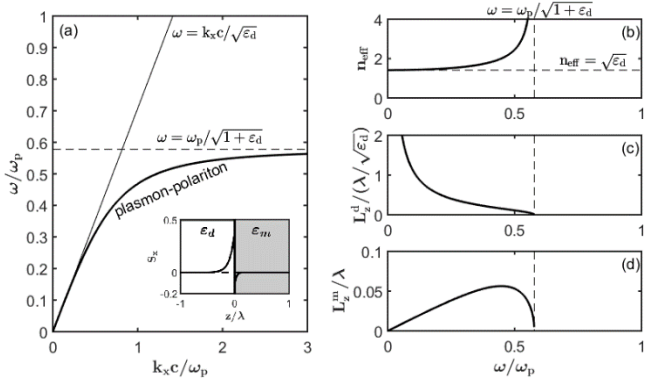

Figure 2. Dispersion relation and fundamental properties of a lossless surface plasmon-polariton ($n_d = 1.33$). (a) Band diagram of a surface plasmon. Insert: $S_x$ energy flux of a plasmon at $\omega = 0.4\omega_p$. Optical properties of a plasmon: (b) Effective refractive index, (c) Penetration depth into dielectric, (d) Penetration depth into metal.

In figure 3 we present similar results for a surface phonon-polariton, and arrive to a similar conclusion that somewhat above the resonant frequency $\omega_0$, the surface phonon-polariton extent into polar material is always deeply subwavelength $L_z^m \ll \lambda$, while its extent into dielectric is generally comparable or even smaller than the wavelength of light $L_z^d < \lambda$ at higher frequencies.

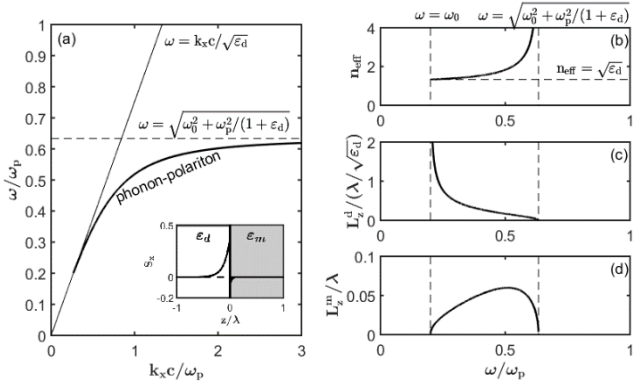

Figure 3. Dispersion relation and fundamental properties of a lossless surface phonon-polariton ($n_d = 1.33, \omega_0 = 0.2\omega_p$). (a) Band diagram of a surface phonon-polariton. Insert: $S_x$ energy flux of a phonon-polariton at $\omega = 0.4\omega_p$. Optical properties of a phonon-polariton: (b) Effective refractive index, (c) Penetration depth into dielectric, (d) Penetration depth into polar medium.

## Lossy materials, case of simple metals at THz frequencies

When material losses are small:

$$\text{Metals: } \gamma \ll \frac{\omega_p}{\sqrt{1+\varepsilon_d}} \quad ; \quad \text{Polar materials: } \gamma \ll \sqrt{\omega_0^2 + \frac{\omega_p^2}{1+\varepsilon_d}} - \omega_0, \qquad (14)$$

conclusions drawn in the case of lossless materials will still hold virtually unchanged, while corrections to the modal optical properties could be found using perturbative expansions with respect to the imaginary part of the dielectric constant. In some cases, however, material losses cannot be ignored as imaginary part of the material dielectric constant can be comparable or larger than its real part. This is notably the case of Drude metals at very low frequencies $\omega \ll \gamma$ operating in the THz and microwave spectral ranges. At such frequencies, metal dielectric constant is essentially purely imaginary:

$$\varepsilon_m(\omega) \approx i \frac{\omega_p^2}{\gamma \omega} \qquad (15)$$

Remarkably, even at very low frequencies a well define surface state with complex dispersion (5) still exists. By substituting (15) into (5) the following expression for the surface plasmon-polariton dispersion relation in the THz range can be found:

$$n_{eff} \approx n_d + \frac{n_d^3}{2}\left(\frac{\omega}{\omega_p}\right)^2 + i \cdot \frac{n_d^3 \omega \gamma}{2\omega_p^2} \qquad (16)$$

As an example, in figure 4 we plot properties of a surface plasmon-polariton propagating at the gold/air interface and observe that while the surface state shows very low loss (long propagation distances) $L_x \gg \lambda$, at the same time, it is highly delocalized into the dielectric material $L_z^m \gg \lambda$. Here we used for gold $\omega_p = 2\pi \cdot 2175\ THz$ and $\gamma = 2\pi \cdot 6.48\ THz$ [12].

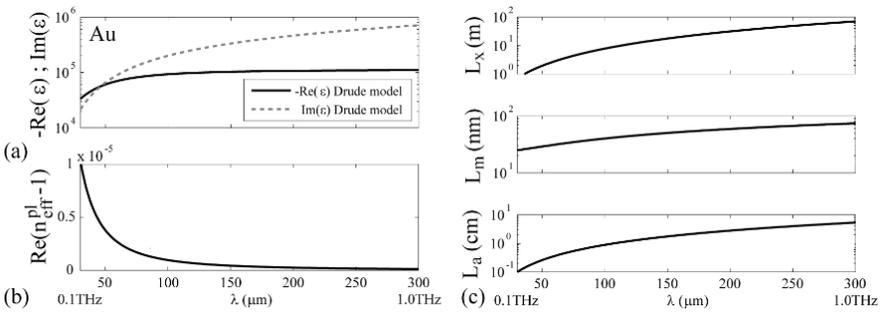

Figure 4. Far-IR spectral range: plasmon propagating at the gold / air interface. (a) Dielectric constant of gold as a function of wavelength. (b) Real part of the plasmon effective refractive index as a function of wavelength. (c) Plasmon propagation length and plasmon penetration depths into dielectric and metal regions.

### Lossy high-k dielectrics

A lossy surface-bound E&M state (also known as a Zenneck wave) can be excited at the interface between two distinct dielectrics both having positive real parts of the dielectric constants. In what follows we assume that one dielectric is lossless $\varepsilon_d > 0$, while another is a high-k lossy dielectric $\varepsilon_m = \varepsilon_m' + i\varepsilon_m''$, where $\varepsilon_m \gg \varepsilon_d$. One can then use directly expressions for the modal effective refractive index (5) and modal extents into the dielectric materials (7), while choosing the signs of the transverse wavevectors in both media to satisfy the surface-bounded wave condition (6). Resultant expressions can be further simplified in the limit $\text{Im}(\varepsilon_m) \ll \text{Re}(\varepsilon_m)$, which is the case for most high-k dielectrics to obtain the following expressions for the modal propagation distance (limited by material losses), and modal extents into the two dielectrics:

$$L_d \approx \frac{\lambda}{\pi} \frac{\sqrt{\varepsilon_d}}{\varepsilon_m''} \left(1 + \frac{\varepsilon_m'}{\varepsilon_d}\right)^{\frac{3}{2}} \quad (17)$$

$$L_m \approx \frac{L_d}{2 + \varepsilon_m'/\varepsilon_d} \quad (18)$$

$$L_x \approx L_d \sqrt{\frac{\varepsilon_m'}{\varepsilon_d}} \quad (19)$$

From these expressions we note that in the case of high-k dielectrics $\varepsilon_m' \gg \varepsilon_d$, Zenneck wave is well defined as its propagation distance is much larger than the modal extents into both dielectrics $L_x \gg L_{m,d}$. At the same time, modal extent into the high-k dielectric is much smaller than that into the lossless dielectric $L_m \ll L_d$. At the same time, achieving strong confinement of Zenneck waves near the surface is somewhat problematic due to inverse dependence of its transverse size on the loss of a high-k dielectric. Nevertheless, as it follows from the further analysis of (17) and (5,7) by using high-k dielectrics with high materials losses $\text{Im}(\varepsilon_m) \sim \text{Re}(\varepsilon_m)$ allows modal sizes in the $L_d \sim 2 - 10 \cdot \lambda$ range, thus making lossy high-k dielectrics viable candidates for surface wave excitation.

## ENABLING MATERIALS REVIEW

In what follows we review several classes of materials beyond simple Drude metals and study their potential for guiding strongly localized surface modes in the Terahertz frequency range. In this work we concentrate mostly on the uniform homogeneous materials which are the simplest to process and shape into desired devices, while only mentioning artificially structured materials (metamaterials) in passing due to complexity of their fabrication.

### Semiconductors

First, we consider semiconductor materials which are directly analogous to metals and can often be described using a simple Drude model (8). However, their plasma frequency generally expressed as:

$$\omega_p = \sqrt{\frac{e^2 N}{\varepsilon_0 m^*}} \tag{20}$$

is significantly lower than that of the metals due to lower concentrations of the free carriers $N$. Ideally, we look for materials that have plasma frequency as close to the THz spectral range as possible, while also having relatively small absorption loss that satisfy (14). Under these conditions, such materials will support tightly localised surface states in THz that are directly analogous to plasmon-polaritons in the visible spectral range.

As an example, the InSb which can support highly confined surface plasmon mode has a plasma frequency of 7.32 THz [11]. Moreover, semiconductor conductivity proprieties can be tuned via adjustment of the free carrier concentrations using temperature or doping. Also, magnetic fields can be used to modulate electrical and optical proprieties of magneto-optical materials [12].

The dependency of the dielectric function of semiconductors on the carrier concentration, effective mass and the damping coefficient, together with an ability of adjustment of their properties via design, make semiconductors promising materials for surface wave excitation in THz spectral range. To illustrate these relations, we present in table I parameters of a n-type and a p-type GaAs films as reported [13].

Table I: Drude parameters for *n*-type and *p*-type GaAs films.

| GaAs | $m^*$ ($m_0$) | $N$ (cm$^{-3}$) | $\tau$ (s) | $\omega_p/2\pi$ (THz) | $\gamma/2\pi$ (THz) |
|---|---|---|---|---|---|
| *n*-type | 0.079 | $4.48 \times 10^{18}$ | $8.47 \times 10^{-14}$ | 67.62 | 1.879 |
| *p*-type | 0.34 | $1.15 \times 10^{18}$ | $2.51 \times 10^{-14}$ | 16.51 | 6.341 |

In figure 5 we plot the complex dielectric function for the *n*-type and *p*-type GaAs films along with the expected plasmon propagation length and penetration depths into air and semiconductor. We observe that the penetration depth of the electric field in air is smaller for the *p*-type (which has plasma frequency closer to Terahertz spectral range) than for the *n*-type. Thus, in this particular example, the *p*-type is a more suitable material for the design of devices where strong confinement of the electric field is required.

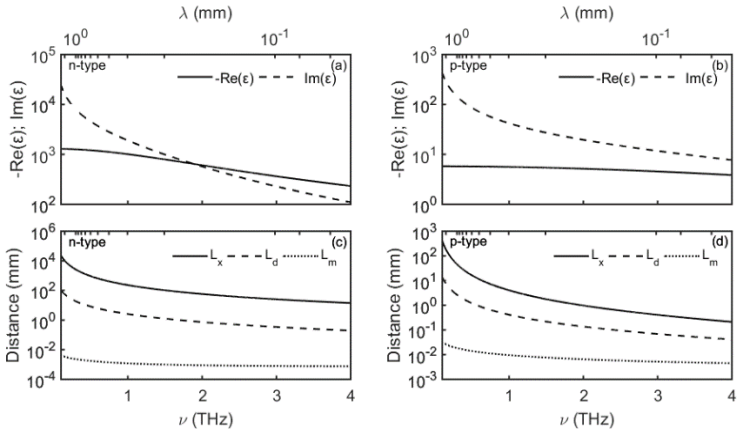

Figure 5. Dielectric function of (a) *n*-type and (b) *p*-type GaAs films following Drude model. Propagation length of the mode $L_x$ and its penetration depth into air $L_d$ and GaAs $L_m$ for (c) *n*-type and (d) *p*-type.

## Polar materials

Next, we consider polar materials, which are characterised by the complex dielectric permittivity that depend also on the intermolecular and intramolecular resonance frequencies. The Lorentz model is typically used to model the frequency response of the dielectric function of polar materials [14]:

$$\tilde{\varepsilon}(\omega) = \varepsilon_\infty \left[1 - \frac{\omega_p^2}{\omega^2 - \omega_0^2 + i\gamma\omega}\right], \quad (21)$$

where $\varepsilon_\infty$ is the dielectric constant at high frequency, $\omega_p$ is the plasma frequency, $\omega_0$ is a resonance frequency and $\gamma$ is the damping coefficient. The real and imaginary parts of the dielectric constant (21) can be expressed as:

$$Re(\tilde{\varepsilon}(\omega)) = \varepsilon_\infty \left[1 - \frac{\omega_p^2(\omega^2 - \omega_0^2)}{(\omega^2 - \omega_0^2)^2 + \gamma^2\omega^2}\right] \quad (22)$$

$$Im(\tilde{\varepsilon}(\omega)) = \frac{\varepsilon_\infty \omega_p^2 \gamma \omega}{(\omega^2 - \omega_0^2)^2 + \gamma^2\omega^2} \quad (23)$$

For such polar materials, the real part of the permittivity can be negative near the resonance frequency $\omega > \omega_0$. As the nature of the resonance, especially at THz frequencies, is often related to the vibrational movements of macromolecules or material domains, then resultant surface waves can be called phonon-polaritons. The table II summarizes Lorentz parameters for some polar materials:

Table II: Lorentz parameters for various polar materials.

| Material | $\varepsilon_\infty$ | $\omega_0/2\pi$ (THz) | $\omega_p/2\pi$ (THz) | $\gamma/2\pi$ (THz) |
|---|---|---|---|---|
| PVDF [15] | 2.0 | 0.3 | 1.47 | 0.1 |
| D-Glucose [16] | 2.940 | 1.446 | 0.037 | 0.021 |
| SrTiO$_3$ [17] | 8.528 | 2.632 | 15.77 | 0.6 |
| LiNbO$_3$ (ordinary) [18] | 13.2 | 4.53 | 6.99 | 0.51 |
| LiNbO$_3$ (extraordinary) [18] | 20.6 | 3.87 | 2.04 | 0.69 |
| LiTaO$_3$ (ordinary) [18] | 13.4 | 4.23 | 6.12 | 0.15 |
| LiTaO$_3$ (extraordinary) [18] | 8.4 | 5.96 | 11.63 | 0.34 |

In figure 6 we show the dielectric properties of the PVDF along with the expected propagation distance and penetration depths in the air and the material. We observe that the modal size in the dielectric is smaller for frequencies closer to the upper limit operation frequency where phonon-polariton can be excited. We also note that at frequencies right below the resonance, polar material dielectric constant can have very large and positive values of both their real and imaginary parts, which is similar to the case of lossy high-k dielectrics. Therefore, right below the resonance, polar materials support Zenneck waves, while above the resonance they support phonon-polaritons.

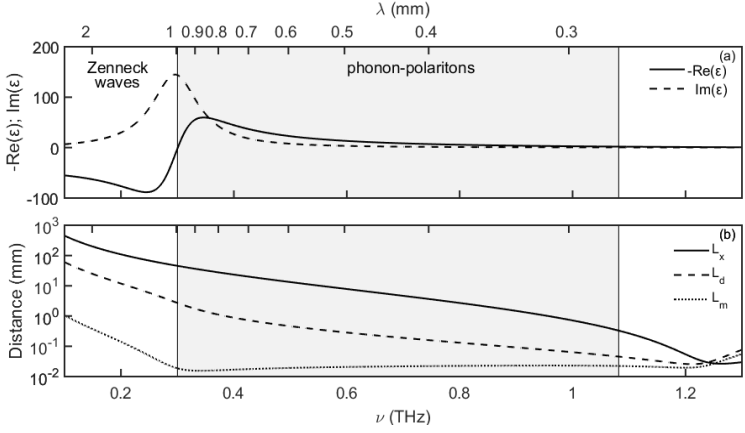

Figure 6. (a) Dielectric function of PVDF following Lorentz model. (b) Propagation length of the mode $L_x$ and its penetration depths into air $L_d$ and PVDF $L_m$.

## Zero-gap materials

Zero-gap materials are semiconductors where the conduction and valent band edges meet at the Fermi level [19]. Here, the electrons can easily change state to fill empty bands. Therefore, these materials are extremely sensitive to external influences such as pressure and magnetic field. The graphene, which consists of a 2D hexagonal arrangement of carbon atoms, is a great example of a zero-gap semiconductor. This material, which can be tuned by magnetic field and chemical doping, could lead to the development of many new electronic devices [20]. The dielectric function of a monolayer of graphene is given by [21]:

$$\tilde{\varepsilon}(\omega) = 1 - \frac{1}{\pi \hbar^2 \varepsilon_0 t_g} \frac{e^2 \mu}{\omega(\omega + i\tau^{-1})}, \qquad (24)$$

where $t_g = 0.34$ nm is the thickness of the graphene monolayer, $\mu$ is the chemical potential and $\tau$ is the scattering rate. In figure 7 we plot the dielectric properties in the THz spectral range of a doped monolayer graphene film of $\mu = 0.135$ eV and $\tau = 1.35 \times 10^{-13}$ s [22]. In contrast to simple metals, real part of the graphene dielectric constant is larger (in absolute value) that its imaginary part for most of the THz spectral range. The plasmon-polariton confinement increases with frequency and modal size becomes commensurable with THz wavelength at higher frequencies ~3THz. Moreover, the modal size in the monolayer is extremely low, thus resulting in very long plasmon propagation distance.

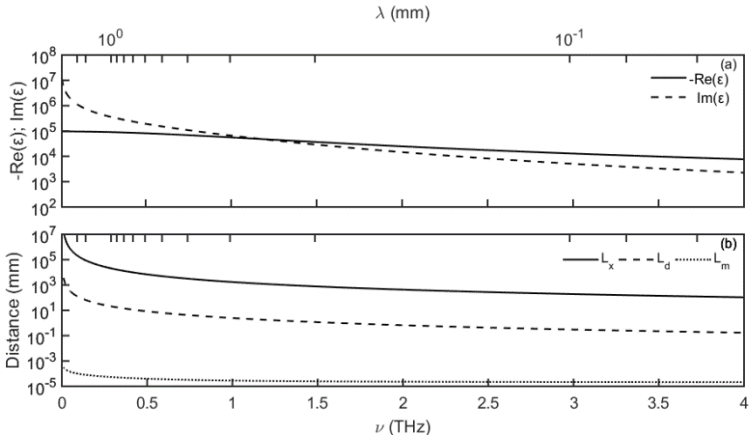

Figure 7. (a) Dielectric function of a doped monolayer of graphene. (b) Propagation length of the mode $L_x$ and its penetration depth into air $L_d$ and graphene $L_m$.

## Lossy high-k dielectrics

As shown in the theory section, at the interface with a lossy high-k dielectric that has imaginary part comparable to its real part, one can excite a well confined lossy surface state. As an example, in figure 8 we present optical properties of the $0.4Ba_{0.6}Sr_{0.4}TiO_3$-$0.6La(Mg_{0.5}Ti_{0.5})O_3$ (BST-LMT) ceramic as reported [23], as well as optical properties of a Zenneck wave that can propagate at the ceramic/air interface. More generally, high-k dielectrics are relatively abundant in the form of various oxides and ceramics, including those that have relatively high imaginary parts of their dielectric constant comparable to their real parts.

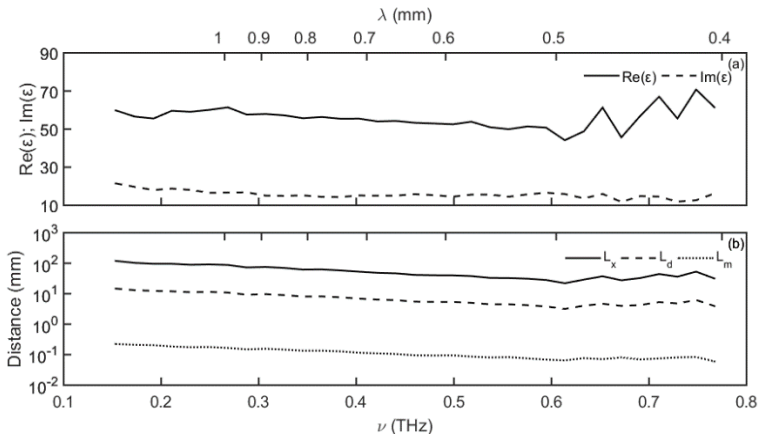

Figure 8. (a) Dielectric function of BST-LMT at room temperature. (b) Propagation length of the surface mode $L_x$, and its penetration depth into air $L_d$ and high-k dielectric $L_m$.

Here, we observe that the modal penetration into BST-LMT is subwavelength at all frequencies $L_m \sim \lambda$. Moreover, despite the relatively large modal extent into air $L_d \sim 10 \cdot \lambda$, the modal propagation length is considerably longer that the modal size $L_x \sim 70 \cdot \lambda \gg L_d$, thus a well-defined surface state can be achieved at the high-k dielectric/air interface.

## Artificially structured materials (metamaterials)

So far, we covered several types of natural homogeneous materials able to support subwavelength modes such as semiconductors, polar materials, zero-gap materials and high permittivity materials. At the same time, highly confined modes in Terahertz range can also be obtained by using structured materials [24]. One type of such structured materials features a periodic array of subwavelength-size grooves inscribed onto a planar metallic slab in air. Thus, for a 1D array of grooves of width $a$, depth $h$ and lattice constant $d$ (period), dispersion relation of the fundamental surface mode supported by such a material (also known as spoof surface plasmons) in the limit $\lambda \gg d$, is given by [25]:

$$\varepsilon_{eff}(\omega) = 1 + \left(\frac{a}{d}\tan\left(\frac{\omega h}{c}\right)\right)^2 \tag{25}$$

while the modal extent into air and its approximation in case of subwavelength $h$ is given by:

$$L_d = \frac{1}{\text{Im}(k_z^d)} = \frac{\lambda}{2\pi}\left(\frac{a}{d}\tan\left(2\pi\frac{h}{\lambda}\right)\right)^{-1} \sim \left(\frac{\lambda}{2\pi}\right)^2 \frac{d}{ah} \tag{26}$$

Note that from (28) it follows that the modal extent into air becomes subwavelength for deeper grooves $h > \frac{d}{a}\frac{\lambda}{(2\pi)^2}$. Figure 9 shows dispersion relations of the fundamental spoof surface plasmon and its size in the air for various groove geometries.

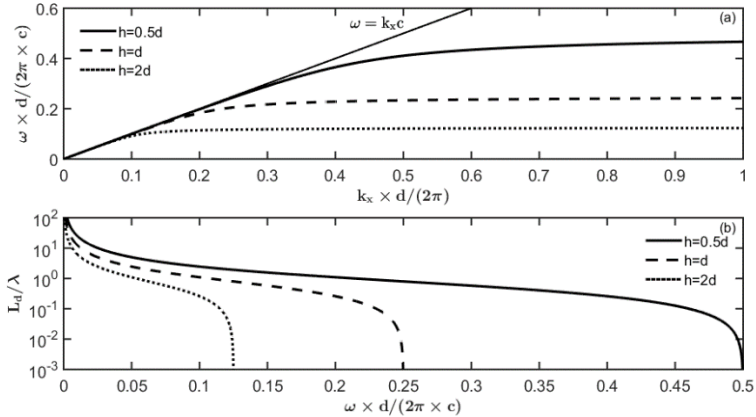

Figure 9. Dispersion relation of a spoof surface plasmon supported by a planar perfect conductor with periodic grooves with $a = 0.2d$. (a) Dispersion relation as a function of the groove depth $h$. (b) Modal penetration depth into air as a function of the groove depth.

At macroscopic scale, this structured material (or metamaterial / effective medium) is analogues to a homogeneous medium able to confine electromagnetic surface waves and can also be interpreted as an artificial metal with a modified plasma frequency [26]. Many other structured materials can be engineered to support propagation of surface plasmon-polaritons-like modes, including 2D patterned and 3D patterned metals [27, 28], helically grooved metal wires [29] as well as dielectric based metamaterials [30]. Moreover, more exotic waves known as Dyakonov waves can be realised at the interface with anisotropic materials which in THz frequency can be, for example, in the form of periodic multilayers of two different materials featuring a deeply subwavelength period (see, for example, [10] for analysis of anisotropic metamaterials in the form of all-dielectric and metal/dielectric planar periodic multilayers with subwavelength period).

## CONCLUSION

In this paper we review several classes of materials such as simple metals, semiconductors, polar materials, zero gap materials, lossy high-k dielectrics as well as structured materials that can support strongly localised electromagnetic surface states in the THz spectral range. We review the basic theory of such waves and present a large number of examples related to naturally occurring and artificial materials. A variety of practical applications is envisioned for such surface waves in THz spectral range including non-destructive super-resolution imaging and quality control, high sensitivity sensors capable of operation with analytes of subwavelength size/thickness, as well as compact optical circuit for the upcoming ultra-high bitrate THz communications.